\documentclass[reprint,aps,pra,showkeys,floatfix,nofootinbib,superscriptaddress]{revtex4-1}

\usepackage{amsmath}
\usepackage{amsfonts}
\usepackage{amssymb}
\usepackage{graphicx}

\numberwithin{equation}{section}

\begin{document}

\title{Robustness of quantum randomness expansion protocols in the presence of noise}

\author{Piotr Mironowicz}
\email{piotr.mironowicz@gmail.com}
\affiliation{Department of Algorithms and System Modelling, Faculty of Electronics, Telecommunications and Informatics, Gda\'{n}sk University of Technology, Gda\'{n}sk 80-233, Poland}
\affiliation{National Quantum Information Centre in Gda\'{n}sk, Sopot 81-824, Poland}

\author{Marcin Paw\l{}owski}
\email{maymp@bristol.ac.uk}
\affiliation{Department of Mathematics, University of Bristol, Bristol BC8 1TW, U.K.}
\affiliation{Institute of Theoretical Physics and Astrophysics, University of Gda\'{n}sk, 80-952 Gda\'{n}sk, Poland}

\date{April 30, 2013}

\begin{abstract}
In this paper we investigate properties of several randomness generation protocols in the device independent framework. Using Bell-type inequalities it is possible to certify that the numbers generated by an untrusted device are indeed random. We present a selection of certificates which guarantee two bits of randomness for each run of the experiment in the noiseless case and require the parties to share a maximally entangled state. To compare them we study their efficiency in the presence of white noise. We find that for different amounts of noise different operators are optimal for certifying  most randomness. Therefore the vendor of the device should use different protocols depending on the amount of noise expected to occur. Another of our results that we find particularly interesting is that using a single Bell operator as a figure of merit is rarely optimal.
\end{abstract}

\maketitle

\section{Introduction}

One of the most striking properties of quantum mechanics is that it is intrinsically random. Moreover, if there exist only a slightly random processes, then also entirely free ones do \cite{GMTDAA12, MP13}.

Random number generation is an important issue in computer science. Random numbers have many applications in such topics as cryptography, authentication \cite{NIST800632}, gambling and system modeling.

Most of the random number generators (RNGs) are basing on purely algebraical manipulation on initial seed. Since the series of numbers produced by such generators are created in a deterministic manner, these RNGs are called pseudo--random number generators (PRNGs). There also exist RNGs basing on some chaotic classical physical processes such as electric or atmospheric noise or by estimating the entropy of hardware interrupts (for example in $/dev/random$ RNG on Linux systems). To make sure that given source of numbers is reliable, some statistical tests may be applied \cite{NIST80022}. Still, these tests can never give a guarantee that the numbers were indeed trustworthy, that is they cannot be predicted by an adversary.

In this situation it is natural to try to use properties of quantum mechanics in order to generate entirely random sequences of numbers. There were efforts making use of such quantum processes like nuclear decay (for example HotBits \cite{HotBits}) or photons hitting a semi--transparent mirror (for example id Quantique RNGs \cite{IDQ}). However when using one of the commercially available quantum random number generators (QRNGs) we still have to trust the vendor of the device.

Therefore many efforts have been made in the quantum information theory to attain a reliability while not trusting the device and even not knowing how does it work. Such an approach, introduced in \cite{DI}, is called device independent. Instead of investigating the internal working of the device in some case it is sufficient to perform tests on its outputs.

Recently the violation of certain Bell inequalities as a certificate of randomness for series from RNG has been used within the device independent approach\cite{ColPHD, RNGCBT, CK11, HWL12, HWL13}. These protocols were randomness expanders, as they use some initial amount randomness to obtain more more of it. In \cite{ColPHD, RNGCBT} as a certificate of randomness the violation of the CHSH Bell inequality\cite{CHSH} was used, while in \cite{CK11} the GHZ correlations were used instead.

\subsection{Purpose of this paper}

Suppose there is a honest vendor that wants to produce and sell QRNGs. His problem is the lack of trust among his potential customers. Since he does not wants to cheat his clients, he can make the design of his device open. But still some parties may distrust that the device is construed in declared manner.

The laws of quantum mechanics give him a way to convince his customers, that they do not need to know the internal working of the device to be sure that they get secure randomness. Using some form of Bell inequalities, they may check, after some statistical tests, that the device produces certain amount of randomness, regardless of the way it has been constructed. Therefore, our honest vendor can propose that his costumers use the protocol described in \cite{RNGCBT}. However, he still needs to decide which Bell inequality to use as a certificate\footnote{The choice of the certificate is his only choice. Therefore, in the whole paper we identify a certificate with a protocol and use these two terms interchangeably}. This device will consist of three parts: two measurement apparatuses and a source of entangled states. The vendor's technology limits the quality (purity) of the states that his source can produce. Let's assume that they are Werner states $\rho_w=p \cdot \left| \psi^- \right\rangle \left\langle \psi^- \right| + (1 - p) \cdot \frac{\openone}{4}$, that is singlets with admixture of white noise. The question that we try to answer in this paper is: Which certificate allows to guarantee most randomness for the given quality of the source measured by $p$? Using it will allow our vendor to maximally exploit the source he has.

All Bell inequalities considered in this paper are maximally violated with pure singlet states. In the noisy case values attainable by Bell operators\footnote{We are assuming that Bell operators are linear functions of correlations, which is more strict that general case where Bell operator is a linear combination of probabilities. All Bell operators used in this paper meet this condition.} are multiplied by $p$.

In this paper we first make a short inspection of min--entropy and Bell inequalities. Then we present $6$ operators, that in noiseless situation may certify two bits of randomness. $3$ of them are based on known Bell inequalities, Braunstein--Caves family \cite{BC88}, CHSH \cite{CHSH} and $T3$ \cite{HWL13}. After that we describe a method we used to find other certificates. Then we give $3$ most interesting examples found this way. Using semi--definite programming we compare the robustness of presented certificates. Finally we investigate the potential of using CHSH inequality to improve presented protocols.

\subsection{Min--entropy}

One of commonly used measures of randomness is min--entropy \cite{OperMinEn}, denoted $H_{\infty}$. For given discrete probability distribution $P = {p_1, \ldots, p_n}$ it is defined as
\begin{equation}
	H_{\infty}(P) \equiv - \log_{2} \left( \max_{i} (p_i) \right)
\end{equation}

Note that min--entropy is directly related to the guessing probability of the value of a particular variable with distribution $P$ with strategy when one guesses the most probable result. In the context of guessing cryptographic keys, min--entropy is a measure of the difficulty of guessing the easiest single key in a given distribution of keys \cite{NIST800632}.

Having some string of characters from a source with given min--entropy per character it is possible to \textit{extract} it's randomness, that is create a shorter string with higher min--entropy per character \cite{Trevisan01, DPVR09, TRSS10}. We use min-entropy as the measure of the efficiency of the protocol throughout the paper.

\section{Randomness certification protocols}

In this paper we investigate the applicability for randomness certification of selected Bell operators, described below. Since we are working in device independent scenario, it is worth to notice that we have no insight into the workings of the device. We do not know what how Alice's and Bob's measurements are carried out, which results may even have outcomes predetermined by the constructor of the apparatus. The only thing we have access to, are outcomes that Alice and Bob get.

We assume that Alice's and Bob's devices are separated during the measurements. This assumption is essential, since only in this case the violation of Bell inequality has any meaning\footnote{In fact this is a way to assert in the device independent scenario that Alice's and Bob's measurements may be treated algebraically as commuting.}. The separation may be for example space--like, if we assume that the signal may not travel between Alice's and Bob's part before the results are collected.

If we want the randomness not only to be fair (for example for gambling and system modeling purposes), but also to be confidential (for example for cryptography or authentication), we also have to assume that the untrusted device do not communicate with the world outside. Without this assumption even the fair RNG may send the results to the adversary.

We will measure the randomness of a given pair of settings using min--entropy. For each pair $(a, b)$ of choices of Alice's and Bob's measurement settings, there exist a distribution $P(A, B|a,b)$ of pairs of outcomes. Min--entropy of the pair of $(a, b)$ is the min--entropy of the distribution $P(A,B|a, b)$.

In the following we assume that all measurements give results $+1$ or $-1$.

We denote by $A_{+}, A_{-}$ Alice's projector on results $+1$ and $-1$ respectively, and similarly $B_{+}, B_{-}$ for Bob. Since outcomes are binary we have $A_{+} + A_{-} = \openone$, and the same for Bob's projectors. Then $C(a, b) = 4 \cdot A_{+} B_{+} - 2 \cdot A_{+} - 2 \cdot B_{+} + \openone$ is the correlation operator. We denote by $Cor(a, b) \equiv P(A = 1, B = 1|a,b) + P(A = -1, B = -1|a,b) - P(A = 1, B = -1|a,b) - P(A = -1, B = 1|a,b)$ the correlations between the binary results obtained by Alice when she chooses measurement $a$ with Bob's results with measurement settings set to $b$. If $\rho$ is the state describing the whole device (including Alice's and Bob's parts, which may be entangled), then $Tr\left(\rho C(a, b) \right)= Cor(a, b)$ and $Cor(a, b)$ may be estimated by collecting statistics of subsequent measurement results.

The measurement settings and the device have to be independent. One of the ways to assure that is to choose these settings randomly. Since in such a situation initial randomness is needed, protocols described below are randomness expanders.

From theoretical point of view it is important to mention the possibility, that from fundamental point of view both measurement choices and results are predetermined. This loophole in Bell inequalities is called superdeterminism and makes all efforts towards generation of randomness pointless.

Below we present $6$ different randomness certification protocols. First of them is based on a well known Braunsein--Caves Bell inequalities. Second one makes use of a pair of Bell operators, which are a decomposition of CHSH inequality. Third one consists of three Bell inequalities, with two of them being CHSH. Three remaining protocols make use of other Bell inequalities described below.

\subsection{Braunstein--Caves inequalities}
\label{prot:BC}

In \cite{BC88} a family of chained Bell inequalities was introduced. The general formula for $n^{th}$ Braunstein--Caves operator is
\begin{equation}
\label{BCn}
	\begin{split}
		BC_{n} & = C(1, 1) + C(1, 2) + C(2, 2) + C(2, 3) + C(3, 3) \\
		& + C(3, 4) + \ldots + C(n-1, n-1) + C(n-1, n) \\
		& + C(n, n)- C(n, 1)
	\end{split}
\end{equation}

The maximal value obtainable in quantum mechanics for $n^{th}$ Braunstein--Caves inequality is $2 \cdot n \cdot \cos \left( \frac{\pi}{2 \cdot n} \right)$. In particular
\begin{equation}
\label{BC3}
	BC_{3} = C(1, 1) + C(1, 2) + C(2, 2) + C(2, 3) + C(3, 3) - C(3, 1)
\end{equation}
is limited by $5.19$.

If we consider the source's quality measured by $p$, then the maximal value is multiplied by $p$.

We check the min--entropy guarantied by the violation of Braunstein--Caves inequality for the following three cases:
\begin{itemize}
	\item for $n = 3$ for the min-entropy of $1^{st}$ setting for Alice and $3^{rd}$ setting for Bob\footnote{Similar results would be obtained for these pairs of settings: $(2, 1)$ and $(3, 2)$.},
	\item for $n = 5$ of settings $(1, 4)$\footnote{Similar results are for pairs $(2, 5)$, $(3, 1)$, $(4, 2)$ and $(5, 3)$.},
	\item for $n = 7$ of settings $(1, 5)$.
\end{itemize}

\subsection{$E_{0}$ and $E_{1}$}
\label{prot:E0E1}
In this case instead of taking only a single operator corresponding to some Bell inequality, we use more than one for randomness certification.

Single CHSH operator may be decomposed into other Bell operators. Let us consider the following two operators:
\begin{subequations}
\label{Es}
 \begin{align}
  E_{0} & = C(1, 1) + C(1, 2) \label{E0} \\
  E_{1} & = C(2, 1) - C(2, 2) \label{E1}
 \end{align}
\end{subequations}

From Uffink's inequality\cite{Uffinka} it follows that the maximal values of \ref{Es} lie on a circle of radius $2$. Taking into account symmetries of these operators, their maximal values obtainable in quantum mechanics may be parameterized in the following manner:
\begin{subequations}
\label{EsMax}
 \begin{align}
  E_{0, max}(\phi) & = 2 \cdot \cos(\phi) \label{E0Max} \\
  E_{1, max}(\phi) & = 2 \cdot \sin(\phi) \label{E1Max}
 \end{align}
\end{subequations}
with $\phi \in \left[0, \frac{\pi}{2}\right]$. The classical limit for the sum of values of these operators is $2$. Thus for $\phi \in \left\{0, \frac{\pi}{2}\right\}$ classical and quantum limits are equal.

Similarly like for previously described Bell operators, if the noise $p$ occurs, then the values in \ref{EsMax} have to be multiplied by this value. In such cases the quantum limit of the sum of \ref{Es} may be achieved for a wider range of values of $\phi$.

Later in this paper we will examine the min--entropy for measurement settings pair $(2, 1)$ as a function of $\phi$ for maximal values of the operators \ref{Es} for different values of noise. Then we will find the optimal angle $\phi$ as a function of noise.

Let us note that this protocolcertificate requires both Alice's and Bob's parts of the device to have only $2$ possible measurement settings. It is the smallest requirements among all presented protocols.

\subsection{$T_3$ with an additional condition}
\label{prot:T3C}

Let us consider a scenario in which Alice has $4$ possible measurement settings and Bob has $3$, each having $2$ possible outcomes. In \cite{HWL13} the following Bell operator was used:
\begin{equation}
	\begin{split}
			T_{3} & = C(1, 1) + C(2, 1) + C(3, 1) + C(4, 1) \\
			& + C(1, 2) + C(2, 2) - C(3, 2) - C(4, 2) \\
			& + C(1, 3) - C(2, 3) + C(3, 3) - C(4, 3) \label{T3}
	\end{split}
\end{equation}

Now let us take two additional Bell operators, which are identical to CHSH with certain choices of settings:
\begin{subequations}
\label{CHSHs}
 \begin{align}
  CHSH_{1} & = C(1, 1) + C(3, 1) + C(1, 2) - C(3, 2) \label{CHSH1} \\
  CHSH_{2} & = C(2, 1) + C(4, 1) + C(2, 2) - C(4, 2) \label{CHSH2}
 \end{align}
\end{subequations}

The maximal value of \ref{T3} that may be obtained in quantum mechanics is $4 \cdot \sqrt{3} \approx 5.928$, and for \ref{CHSHs} is $2 \cdot \sqrt{2} \approx 2.82$.

If we impose on the device a condition that both operators \ref{CHSHs} achieve the value of at least $0 \leq C \leq 2 \cdot \sqrt{2}$, then the maximal value of \ref{T3} is a function of $C$, $T_{3, max} = T_{3, max}(C)$. We require the device to obtain this maximal value. In the case when we cope with noise or imperfections of the device, where $0 < p < 1$, then we have to multiply all \ref{T3} and \ref{CHSHs} by $p$.

Assuming these conditions we will check the lower bound on the min--entropy, as a function of $C$ with maximal possible value of \ref{T3}, for Alice's setting $1$ and Bob's $3$.

\subsection{Certificates obtained randomly}
\label{prot:random}

The three previous cases were chosen by us because they rely, at least  to some extent on known Bell inequalities. In order to learn more about Bell operators certifying randomness we have used some randomized method of finding them.

Most of the known interesting operators are in form
\begin{equation}
\label{BellOpForm}
	\sum_{i, j} \alpha_{i, j} \cdot C(i, j)
\end{equation}
where $\alpha \in \{-1, 0, 1\}$, $i$ enumerates Alice's settings, and $j$ Bob's settings.

We have considered operators that have this form with $4$ settings for Alice and $3$ settings for Bob. In this case there is around half a million different operators and we have randomly chosen a representative sample of around 25 thousand for further studies. Then, for each choice we have computed min--entropy with noise parameter $p = 0.95$ using semi--definite programming (which method is described in more details in section \ref{sec:results} with results). The histogram presenting the number of operators that certify given randomness is shown in table \ref{tab:histogram}.

\begin{table}[ht]
	\centering
		\begin{tabular}{|r|c|c|c|c|c|c|}
			\hline
			min--entropy & 0-0.05 & 0.05-0.1 & 0.1-0.15 & 0.15-0.2 \\
			Bell inequalities & 12853 & 689 & 722 & 696 \\ \hline
			min--entropy & 0.2-0.25 & 0.25-0.3 & 0.3-0.35 & 0.35-0.4 \\
			Bell inequalities & 939 & 907 & 850 & 1324 \\ \hline
			min--entropy & 0.4-0.45 & 0.45-0.5 & 0.5-0.55 & 0.55-0.6 \\
			Bell inequalities & 1176 & 839 & 1062 & 678 \\ \hline
			min--entropy & 0.6-0.65 & 0.65-0.7 & 0.7-0.75 & 0.75-0.8 \\
			Bell inequalities & 493 & 155 & 91 & 15 \\ \hline
		\end{tabular}
	\caption{Number of Bell inequalities that have been randomly chosen depending on the min--entropy they certify under noise $p = 0.95$.}
	\label{tab:histogram}
\end{table}

The most interesting operators are these, which certify the most randomness.

Among tested operators $41$ certified more than $0.72$ bits of randomness under high noise. These operators form $4$ distinct groups that have identical maximal value and randomness certification properties. First of these groups (with $9$ drawn instances) of them revealed to be isomorphic\footnote{Up to two operations. First is a reordering of measurement settings, and second is a change of signs of results for one setting of one party.} to $BC_{3}$ described in \ref{prot:BC}. The remaining three groups are described in \ref{prot:modCHSH} (with $6$ instances drawn) and \ref{prot:other} (with $15$ and $11$ instances drawn, respectively).

\subsubsection{Modified CHSH}
\label{prot:modCHSH}

Now let us consider the following Bell operator which is similar to one used in \cite{DimWit}. We call in modified CHSH because it is a CHSH operator with one additional correlation function.

Let us take operator of the following form:
\begin{equation}
\label{modCHSH}
	C(1, 2) + C(1, 3) + C(2, 1) + C(2, 2) - C(2, 3)
\end{equation}
Under quantum mechanics its value is limited to $1 + 2 \cdot \sqrt{2}$. First four terms form a CHSH operator.

The protocol requires the device to reach the maximal value of this operator (multiplied by $p$ in case with noise).

Further in the paper we will examine the min--entropy with pair of settings $(1, 1)$.

\subsubsection{Other inequalities}
\label{prot:other}

Let us consider the following Bell inequalities:
\begin{equation}
\label{I1}
 \begin{split}
  I_{1} = C(1, 2) - C(1,3) - C(2, 1) - C(2, 2) \\
		+ C(3, 1) + C(3, 3) + C(4, 1) \\
		\leq 1 + 6 \cdot \cos \left( \frac{\pi}{6} \right) \approx 6.19
 \end{split}
\end{equation}
and
\begin{equation}
\label{I2}
 \begin{split}
	I_{2} = -C(1,2) + C(1, 3) + C(2, 1) + C(2, 2) + C(2, 3) \\
		+ C(3, 2) - C(3, 3) + C(4, 1) + C(4, 2) + C(4, 3) \\
		\leq 2 + 4 \cdot \sqrt{2} \approx 7.66
 \end{split}
\end{equation}

These two were taken as examples from wider groups of inequalities, that uses $4$ measurement settings of Alice and $3$ of Bob, and consist of $7$, respectively $10$, correlations. Inequalities in both of these groups have the same efficiency in generating min--entropy with noise. Inequalities form the first group are similar to these from Braunstein--Caves family. The pair of measurement settings for which the min--entropy will be investigated is $(1, 1)$.

The same as in protocols based on Braunstein--Caves inequalies (\ref{prot:BC}) and modified CHSH (\ref{prot:modCHSH}), random number generation protocols using these inequalities require the device to reach the maximal value of appropriate operator (multiplied by $p$ in case of noise).

\section{Improving certificates with the CHSH inequality}
\label{sec:CHSH}

Although CHSH inequality is not able to certify two bits of randomness even for $p = 1$, it is quite efficient for $p \leq 0.9$. In fact in this case it is able to guarantee more randomness than most of the two bit protocols, as it is shown in the table \ref{tab:chsh}.

\begin{table}[ht]
	\centering
		\begin{tabular}{|r|c|c|}
			\hline
			p & global & local \\ \hline
			$0.99999$ & $1.21757$ & $0.99090$ \\ \hline
			$0.999$ & $1.12231$ & $0.91155$ \\ \hline
			$0.95$ & $0.58411$ & $0.47234$ \\ \hline
			$0.9$ & $0.37757$ & $0.30718$ \\ \hline
			$0.8$ & $0.13510$ & $0.11362$ \\ \hline
		\end{tabular}
	\caption{Randomness certified by CHSH inequality for different noises (see text).}
	\label{tab:chsh}
\end{table}

Knowing this property some of the above certificates may be improved if an additional condition for CHSH inequality is imposed. This can be only done when the original certifying operator includes CHSH which is the case in (\ref{T3}) and (\ref{modCHSH}).

In the table \ref{tab:T3vsT3C} certificate \ref{prot:T3C} (T3C) is compared with it's version taking into the account only the maximal violation of (\ref{T3}) attainable for the given amount of noise.

\begin{table}[ht]
	\centering
		\begin{tabular}{|r|c|c|}
			\hline
			p & $T3$ & $T3C$ \\ \hline
			$0.99999$ & $1.3294$ & $1.7871$ \\ \hline
			$0.999$ & $1.2171$ & $1.4101$ \\ \hline
			$0.95$ & $0.55873$ & $0.5931$ \\ \hline
			$0.9$ & $0.19515$ & $0.3072$ \\ \hline
			$0.8$ & $0$ & $0.1136$ \\ \hline
		\end{tabular}
	\caption{Comparison of min--entropies certified by $T3$ (\ref{T3}) alone and with two additional CHSH conditions (\ref{prot:T3C}).}
	\label{tab:T3vsT3C}
\end{table}

CHSH inequality has also demonstrated its effectiveness for improving certificate \ref{prot:modCHSH} (modified CHSH). Combining it with a condition $C(1, 2) + C(1, 3) + C(2, 2) - C(2, 3) \geq p \cdot 2 \cdot \sqrt{2}$ gives the results shown in the table \ref{tab:modCHSHPlus}.

\begin{table}[ht]
	\centering
		\begin{tabular}{|r|c|c|}
			\hline
			p & Original \ref{prot:modCHSH} & Improved \ref{prot:modCHSH} \\ \hline
			$0.99999$ & $1.9764$ & $1.9764$ \\ \hline
			$0.999$ & $1.7751$ & $1.7751$ \\ \hline
			$0.95$ & $0.7775$ & $0.78024$ \\ \hline
			$0.9$ & $0.4365$ & $0.45443$ \\ \hline
			$0.8$ & $0.0468$ & $0.1342$ \\ \hline
		\end{tabular}
	\caption{Certificate \ref{prot:modCHSH} with additional CHSH condition compared to the original one.}
	\label{tab:modCHSHPlus}
\end{table}

This protocol for low noises is almost as efficient as \ref{prot:BC} ($BC_{3}$, which is the most efficient in this case), while requiring less measurement settings. For high noises this protocol is very close to \ref{prot:E0E1} ($E_{0}$ with $E_{1}$), while not requiring to fit parameters for particular $p$ parameter. For intermediate amounts of noise it's the best one.

CHSH appears also in (\ref{I2}) but in this case imposing its violation does not help. We believe that it stems from the fact that (\ref{I2}) and any CHSH operator appearing in ti cannot be simultaneously maximally violated.

\section{Results}
\label{sec:results}

The following results were obtained using the \textit{NPA} method introduced and developed in papers \cite{NPA07, NPA08}. This method brings out an infinite hierarchy of conditions, that is satisfied by any set of quantum correlations. Each level of this hierarchy may be mapped into the semi-definite optimization problem. Such a problem may be efficiently solved numerically using the primal--dual interior point algorithm \cite{SeDuMi, SeDuMi102, IntPoint}.

All the results were obtained using level $Q_{2}$ of the NPA hierarchy, only results for Braunstein--Caves inequality for $n = 7$ was calculated in level $Q_{1+AB}$ due to high computer's memory consumption. Since the conditions of these levels do not contain all the laws of quantum mechanics, when computing min--entropy we get a \textit{lower} bound of its value, so all of the examined protocols may give even more randomness than the presented data shows\footnote{In other words, we assume less than quantum mechanics. In particular we do not assume no--signaling principle.}. In fact each of the levels of the hierarchy corresponds to some set of polynomial conditions of a finite degree. In all the cases we take into account the worst case, that is maximize each of the probabilities of given output, independently.

\begin{figure}[hp]
	\centering
		\includegraphics[width=0.45\textwidth]{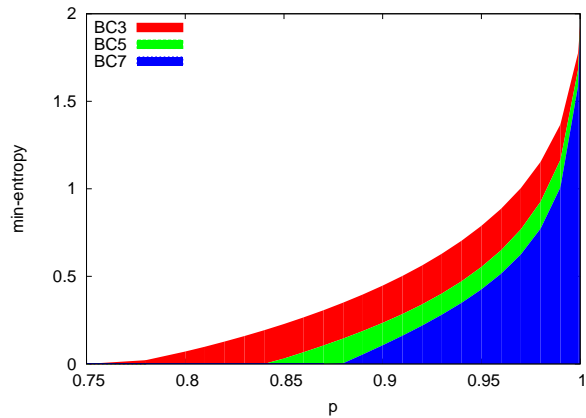}
	\caption{Comparison of lower bounds on min--entropies for protocols described in \ref{prot:BC} (based on Braunstein--Caves inequalities) as a function of noise.}
	\label{fig:BC357}
\end{figure}

The results for the protocol \ref{prot:BC}, that uses Braunstein--Caves operators, as a function of noise, are shown on the figure \ref{fig:BC357}. For all values of noise the simplest operator, $BC_{3}$ (see \ref{BC3}) gives the highest min--entropy.

\begin{figure}[h]
	\centering
		\includegraphics[width=0.45\textwidth]{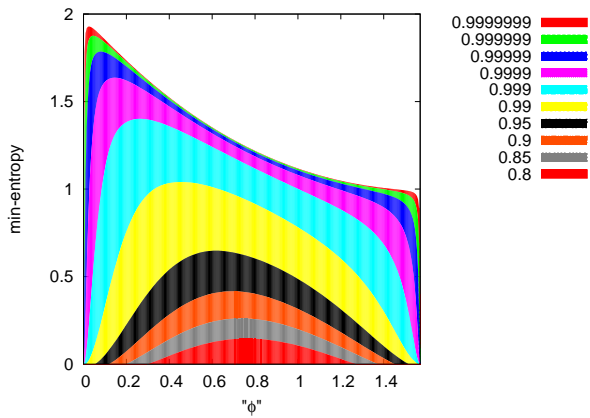}
	\caption{Lower bound on min--entropy for protocol described in \ref{prot:E0E1} ($E_{0}$ with $E_{1}$) as a function of $\phi$ (see equation \ref{EsMax}) for different noises.}
	\label{fig:E0E1}
\end{figure}

In the figure \ref{fig:E0E1} the results for protocol described in \ref{prot:E0E1} ($E_{0}$ with $E_{1}$) are shown. For parameter $\phi$ equal $0$ and $\frac{\pi}{2}$ the min--entropy is $0$, since then the possible values of \ref{Es} in quantum and classical cases are the same, so device's behavior may be implemented classically giving no warranty on randomness. An important result is that the optimal angle between average values of operators \ref{Es} depends on the noise parameter $p$. This dependence is shown on the table \ref{tab:E0E1OptAngle}.

\begin{table}[ht]
	\centering
		\begin{tabular}{|r|l|}
			\hline
			p & $\phi$  \\ \hline
			$0.9999999$ & $0.0252$ \\ \hline
			$0.999999$ & $0.0452$ \\ \hline
			$0.99999$ & $0.0811$ \\ \hline
			$0.9999$ & $0.1460$ \\ \hline
			$0.999$ & $0.2638$ \\ \hline
			$0.99$ & $0.4562$ \\ \hline
			$0.95$ & $0.6179$ \\ \hline
			$0.9$ & $0.6948$ \\ \hline
			$0.85$ & $0.7357$ \\ \hline
			$0.8$ & $0.7617$ \\ \hline
		\end{tabular}
	\caption{Optimal angle between $E_{0}$ and $E_{1}$ depending on noise.}
	\label{tab:E0E1OptAngle}
\end{table}

\begin{figure}[h]
	\centering
		\includegraphics[width=0.4\textwidth]{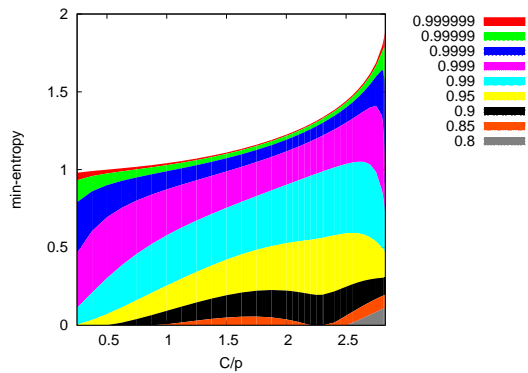}
	\caption{Lower bound on min--entropy for protocol described in \ref{prot:T3C} (T3C) as a function of $C$ (see text) for different noises. }
	\label{fig:T3C}
\end{figure}

The figure \ref{fig:T3C} shows the results for protocol described in \ref{prot:T3C} (T3C). It is worth to notice that as parameter $C$ approaches it maximal value $2 \cdot \sqrt{2}$, then the min--entropy tends to $1$ (in the case without noise).  The min--entropy strongly depends on $C$. This dependence is shown in the table \ref{tab:OptC}.

\begin{table}[ht]
	\centering
		\begin{tabular}{|r|c|c|c|c|}
			\hline
			p & $0.999999$ & $0.99999$ & $0.9999$ & $0.999$ \\ \hline
			$\frac{C}{p}$ & $2.826$ & $2.82$ & $2.8$ & $2.75$ \\ \hline \hline
			p & $0.99$ & $0.95$ & $0.9$ & $0.8$ \\ \hline
			$\frac{C}{p}$ & $2.6$ & $2.55$ & $2.828$ & $2.828$ \\ \hline
		\end{tabular}
	\caption{Optimal value of parameter $C$ for protocol \ref{prot:T3C} (T3C) depending on noise.}
	\label{tab:OptC}
\end{table}

Considering the bound on the value of the operator $T3$ (\ref{T3}, not shown on the figure), it is maximal for $C = 2.3094116$. For high noises ($p < 0.95$) min--entropy has local minimum near this point.

\begin{figure}[hp]
	\centering
		\includegraphics[width=0.4\textwidth]{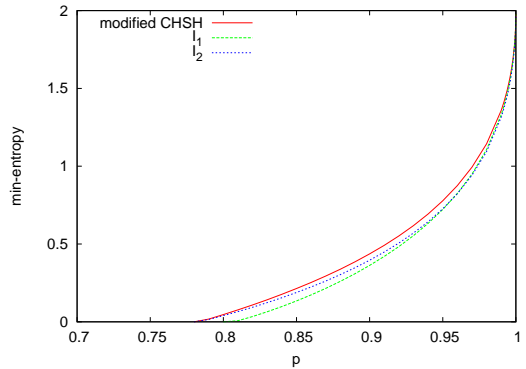}
	\caption{Comparison of lower bounds on min--entropies for protocols described in \ref{prot:modCHSH} (modified CHSH) and \ref{prot:other} (other randomly generated protocols) as a function of noise.}
	\label{fig:random}
\end{figure}

The figure \ref{fig:random} contains the results for protocols using inequalities described in \ref{prot:modCHSH} (modified CHSH) and \ref{prot:other} (other randomly generated operators).

It can be seen that the protocol \ref{prot:BC} using Bell inequality \ref{BC3} (BC3) gives the largest amount of randomness comparing to other described protocols, when the noise parameter $p$ is larger than $0.9$. Different operators from Braunstein--Caves family give less min--entropy for all amounts of noise and requires more settings for each party\footnote{It is important for a protocol to use not many measurement settings, as they require more initial randomness for expansion.}.

For high noises ($p \approx 0.8$) the largest min--entropy is obtained using protocol \ref{prot:E0E1} ($E_{0}$ with $E_{1}$). The main disadvantage of this protocol is the necessity to chose the angle parameter individually for each noise.

The certificate \ref{prot:T3C} (T3C) shares with \ref{prot:E0E1} ($E_{0}$ with $E_{1}$) the need for choosing its parameter (in this case $C$) for the given noise. Similarly it gives good results for high noises ($p \approx 0.8$), but not as good as \ref{prot:E0E1}. It also requires more measurement settings.

Using modified CHSH (\ref{prot:modCHSH}) and certificates $I_{1}$ and $I_{2}$ (\ref{prot:other}) for smaller noises ($p \geq 0.9$), the amount of achieved randomness is slightly smaller that the randomness form the protocol $BC_{3}$ (\ref{prot:BC} with Bell inequality \ref{BC3}). However, protocol based on modified CHSH inequality requires less measurement settings than $BC_{3}$.

Protocols with more complicated Bell inequalities $I_{1}$ and $I_{2}$ slightly differ depending on the amount of noise. For $p \geq 0.999$ the inequality $I_{1}$ (\ref{I1}) gives more randomness than $I_{2}$ (\ref{I2}), while for $p \leq 0.999$ $I_{2}$ gives more randomness than $I_{1}$.

Comparison of all protocols described in this paper may be found in the tables \ref{tab:compare1} and \ref{tab:compare2}.

\begin{table}[ht]
	\centering
		\begin{tabular}{|r|c|c|c|c|c|}
			\hline
			p & $BC_{3}$ & $BC_{5}$ & $BC_{7}$ & $E_{0} E_1$\footnote{Values for optimal angle parameter.} & $T3C$ \\ \hline
			$0.99999$ & $1.9769$ & $1.9656$ & $1.9537$ & $1.7854$ & $1.7871$ \\ \hline
			$0.999$ & $1.7792$ & $1.6841$ & $1.5917$ & $1.4013$ & $1.4101$ \\ \hline
			$0.95$ & $0.7885$ & $0.5534$ & $0.4258$ & $0.6484$ & $0.5931$ \\ \hline
			$0.9$ & $0.4474$ & $0.2342$ & $0.1064$ & $0.4163$ & $0.3072$ \\ \hline
			$0.8$ & $0.0709$ & $0.0000$ & $0.0000$ & $0.1461$ & $0.1136$ \\ \hline
		\end{tabular}
	\caption{Comparison of protocols $BC_{3}$, $BC_{5}$, $BC_{7}$ (\ref{prot:BC}), $E_{0} E_1$ (\ref{prot:E0E1}) and $T3C$ (\ref{prot:T3C}) for different noises.}
	\label{tab:compare1}
\end{table}

\begin{table}[ht]
	\centering
		\begin{tabular}{|r|c|c|c|c|c|}
			\hline
			p & modified CHSH & $I_{1}$ & $I_{2}$ \\ \hline
			$0.99999$ & $1.9764$ & $1.9753$ & $1.9742$ \\ \hline
			$0.999$ & $1.7751$ & $1.7649$ & $1.7558$ \\ \hline
			$0.95$ & $0.78024$ & $0.7219$ & $0.7262$ \\ \hline
			$0.9$ & $0.45443$ & $0.3625$ & $0.3959$ \\ \hline
			$0.8$ & $0.1342$  & $0.0000$ & $0.0398$ \\ \hline
		\end{tabular}
	\caption{Comparison of protocols \ref{prot:modCHSH} (modified CHSH) and \ref{prot:other} ($I_{1}$ and $I_{2}$) for different noises. In the column corresponding to modified CHSH we give the result for its improved version.}
	\label{tab:compare2}
\end{table}

\section{Conclusions}

In this paper we presented several certificates for randomness generation protocols. We find that there is no unequivocally  most efficient one for all the amounts of noise. On the other hand some are clearly better than the others.

For low noises ($p \geq 0.92$) the most randomness may be obtained using the certificate \ref{prot:BC} based on Braunstein--Caves inequality. This protocol requires $3$ binary measurements for Alice and for Bob and uses only one inequality.

In the case of intermediate noise ($0.85 \leq p \leq 0.92$) the protocol \ref{prot:modCHSH} with additional CHSH condition (\ref{sec:CHSH}) certifies most randomness. It requires $2$ binary measurements for Alice and $3$ for Bob.

Protocol \ref{prot:E0E1} based on a pair of operators, $E_{0}$ and $E_{1}$, certifies most randomness among all the compared protocols for high noises ($p \leq 0.85$). It requires $2$ binary measurements for both Alice and Bob and has a parameter $\phi$ that has to be chosen for a particular noise $p$ for optimal results.

\begin{figure}[h]
	\centering
		\includegraphics[width=0.4\textwidth]{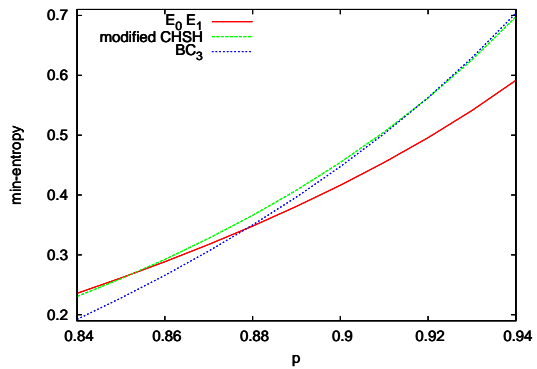}
	\caption{Comparison of three most efficient of investigated protocols for $0.84 \leq p \leq 0.94$.}
	\label{fig:BestChartLarge}
\end{figure}

A comparison of these three protocols is shown on figure \ref{fig:BestChartLarge}.

The conclusion of this paper is that a honest vendor of QRNGs should use one of these three protocols, depending on the amount of noise he is expecting the device will have to cope with.

The first interesting fact that our research has revealed is that there is no single optimal certificate. But what is even more remarkable is that the three best ones fall into three distinct categories. Certificate \ref{prot:BC} is just a Bell operator;  \ref{prot:modCHSH} is a combination of two; and $E_{0}$ and $E_{1}$ are not even Bell inequalities, furthermore they have to be considered with different weights. This proves that there is more than one place to look for optimal certificates and although Bell inequality violation is a necessary condition for device independent certification of randomness it is not always a good measure of it.

\section{Acknowledgements}
SDP was implemented in OCTAVE using SeDuMi \cite{SeDuMi} toolbox. This work is supported by FNP TEAM, IDEAS PLUS and U.K. EPSRC.

\end{document}